\documentclass[10pt,amsmath,amssymb,aps,pra,twocolumn,longbibliography,floatfix,superscriptaddress,nofootinbib]{revtex4-1}

\usepackage[utf8]{inputenc}
\usepackage[T1]{fontenc}

\usepackage{dsfont}
\usepackage{times}
\usepackage{txfonts}

\usepackage{physics}
\usepackage{mathtools}
\usepackage{mleftright}\mleftright
\usepackage{float}
\usepackage{enumitem}
\usepackage{bm}

\newtheorem{definition}{Definition}

\usepackage{algorithm}
\usepackage{algpseudocode}
\floatstyle{boxed}
\newfloat{algorithm}{t}{lop}
\floatname{algorithm}{Algorithm} 

\usepackage{multirow}
\usepackage{diagbox}
\usepackage{soul}
\floatstyle{boxed}
\usepackage{dcolumn}

\usepackage{graphicx}
\usepackage{xcolor}

\usepackage[
final,
colorlinks,
allcolors = blue
]{hyperref}

\begin{document}
\title{Counting collisions in random circuit sampling for benchmarking  quantum computers}
\author{Andrea Mari}
\affiliation{Unitary Fund}
\affiliation{Physics Division, School of Science and Technology, Universit\`a di Camerino, 62032 Camerino, Italy}

\begin{abstract}
We show that counting the number of collisions (re-sampled bitstrings) when measuring a random quantum circuit provides a practical benchmark for the quality of a quantum computer and a quantitative noise characterization method. We analytically estimate the difference in the expected number of collisions when sampling bitstrings from a pure random state and when sampling from the classical uniform distribution. We show that this quantity, if properly normalized, can be used as a {\it collision anomaly} benchmark or as a {\it collision volume} test which is similar to the well-known quantum volume test, with advantages (no classical computing cost) and disadvantages (high sampling cost).
We also propose to count the number of  cross-collisions between two independent quantum computers running the same random circuit in order to obtain a cross-validation test of the two devices.
Finally, we quantify the sampling cost of quantum collision experiments. We find that the sampling cost for running a collision volume test on state-of-the-art processors (e.g.~20 effective clean qubits) is quite small: less than $10^5$ shots. For large-scale experiments in the quantum supremacy regime, the required number of shots for observing a quantum signal in the observed number of collisions is currently infeasible ($>10^{12}$), but not completely out of reach for near-future technology.
\end{abstract}

\maketitle

\section{Introduction}

Validating and benchmarking the quality of quantum computers is an important task \cite{eisert2020quantum, amico2023defining, wang2022sok, resch2021benchmarking},  especially in the present technological era of small-scale and noisy quantum processors \cite{preskill2018quantum, corcoles2019challenges}. In addition to the characterization and testing of quantum processors \cite{emerson2005scalable, magesan2011scalable, bishop2017quantum, moll2018quantum, cross2019validating, blume2020volumetric, baldwin2022re, kechedzhi2023effective, larose2022error, helsen2022general, helsen2023shadow, elben2020cross, lubinski2023application,lubinski2023optimization, mills2021application, mundada2023experimental, li2023qasmbench, helsen2023shadow}, the validation problem is also relevant for large-scale quantum computations beyond the classical simulablity regime \cite{boixo2018characterizing,  mahadev2018classical, aaronson2019classical}, also known as the {\it quantum supremacy} regime \cite{preskill2012quantum, arute2019quantum, wu2021strong, madsen2022quantum, zhong2020quantum}. In fact, one of the most difficult tasks in quantum supremacy experiments is proving that the quantum processor is reliable or, at least, reliable with a nonzero fidelity.

In this work we propose a remarkably simple validation protocol that can be summarized in a single sentence: repeatedly execute a random quantum circuit and count the number of collisions, i.e., the number of times a bitstring, that was previously sampled, is sampled again \cite{footnote1}.
It turns out that the expected number of collisions is larger for a good quantum computer compared to a noisy one, or compared to the extreme limit of a (classical) random number generator. By comparing the observed number of collisions with the theoretical ideal value, one can quantitatively deduce the fidelity of the quantum computation. More practically, the same experimental procedure can be used to obtain a {\it collision volume} (CV) benchmark which, like the standard quantum volume metric \cite{cross2019validating, blume2020volumetric, baldwin2022re}, quantifies the maximum number of clean {\it effective qubits} \cite{bishop2017quantum, moll2018quantum, kechedzhi2023effective, larose2022error} available in the quantum device. 

\begin{figure}[!t] 
    \includegraphics[width=1.0 \linewidth]{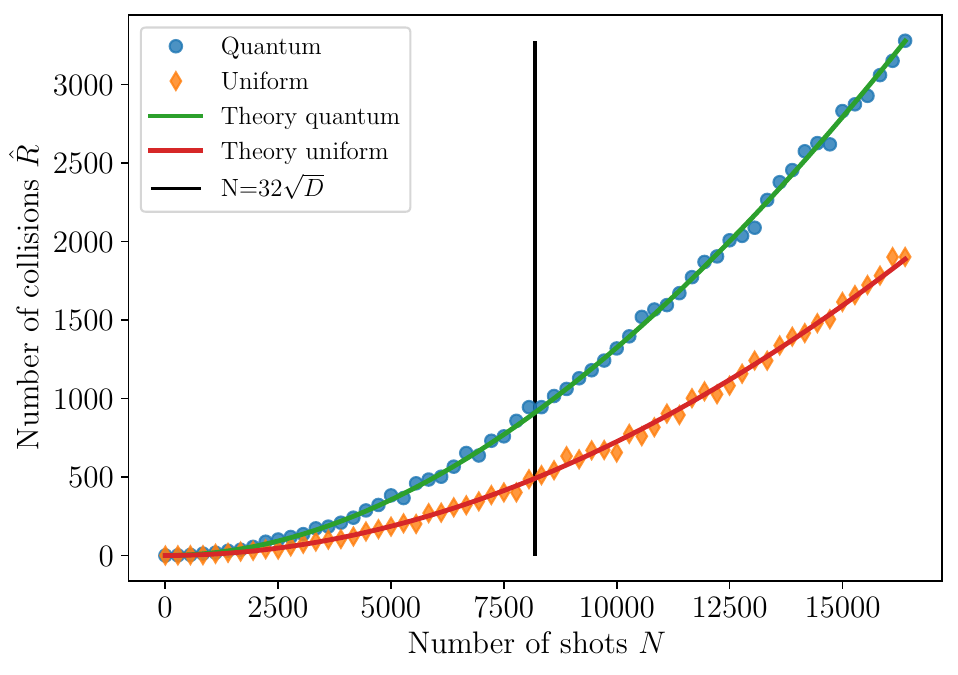}
    \caption{Numerical simulation of the number of collisions $\hat R$ with respect to the number of samples $N$,  for a single $16$-qubit random pure quantum state (blue) and for the uniform distribution over bitstrings of $n$=16 bits (orange). The theoretical lines correspond to Eqs. \eqref{eq:u_collisions} and \eqref{eq:q_collisions}. The vertical line is the educated guess for the number of shots required to observe $\gtrsim 500$ collisions, as derived in Eq.\ \eqref{eq:educated_guess}. $D$ is the dimension of the sample space which, in this example, is $2^{16}$.} \label{fig:collisions}
\end{figure}

Several methods and protocols have been proposed for benchmarking quantum computations \cite{eisert2020quantum, amico2023defining, wang2022sok, resch2021benchmarking}. Such methods can be classified in two main categories: application-oriented benchmarks \cite{lubinski2023application, mills2021application, mundada2023experimental, li2023qasmbench, lubinski2023optimization} designed to quantify performance on specific problems, and randomized benchmarks \cite{emerson2005scalable, magesan2011scalable, bishop2017quantum, moll2018quantum, cross2019validating, blume2020volumetric, baldwin2022re, boixo2018characterizing, aaronson2019classical, kechedzhi2023effective, larose2022error, helsen2022general} designed to measure the overall quality of a quantum computer.
Two widely used randomized methods for validating a quantum device are the quantum volume (QV) test \cite{cross2019validating, blume2020volumetric, baldwin2022re} and the cross-entropy benchmarking (XEB) method \cite{boixo2018characterizing, aaronson2019classical}. In both cases, random quantum circuits are executed and the distributions of the measurement outcomes are compared against the corresponding ideal probability distributions. The initial step of those protocols is equal to the initial step of the method proposed in this work, i.e., collecting samples from random circuits. However, while the QV and XEB approaches require a classical simulator to compute the ideal measurement probabilities, in our quantum collision approach no classical computing is required. In the quantum collision approach a quantum device is benchmarked {\it against itself} or, as we also discuss later, against a different quantum device.

\begin{table*}[!t]
    \centering
    \begin{tabular}{l|c|c|c|c|c}
        Benchmarks based on random circuits & Classical cost    & Quantum cost   &  Incoherent errors       & Coherent errors   & Classical spoofing hardness            \\
        \hline
       QV \cite{blume2020volumetric} and linear XEB  \cite{boixo2018characterizing, aaronson2019classical} &   exponential     &  polynomial     &     sensible             &   sensible                   & Hard \cite{aaronson2019classical} with ongoing debate \cite{barak2020spoofing}    \\
       Collision test (this work)       &      none         &  exponential    &     sensible to most     &   insensible                 & Easy               \\
       Cross-collision test (this work)  &      none         &   exponential   &     sensible             &   sensible                   & Unknown (assuming no cooperation) \\
       \end{tabular} 
    \caption{Qualitative summary of the main properties of collision-based benchmarks and comparison with other existing methods. By {\it sensible to most errors} we mean that the collision test executed on a single device is sensible to most incoherent errors in a real-world scenario, but one can easily find ad-hoc counterexamples (e.g. a noise model that, with some probability, replaces the output state of the circuit with a fixed product state). We remark that if Alice and Bob can agree on a common cheating strategy (e.g. on a secret set of shared bitstrings) the cross-collision test can be trivially spoofed. On the contrary, assuming Alice and Bob cannot adversarially cooperate, we conjecture that the test is as hard to spoof as linear XEB.}
    \label{tab:comparison}
\end{table*}

Unfortunately, there is no free lunch since the exponential classical simulation cost required by the QV and XEB methods is pushed, by the quantum collision protocol, to the quantum side. Indeed, in order to observe a nonzero number of collision events when sampling a random state of $n$ qubits, one needs a number of shots larger than $2^{n/2}$. This is an exponential cost, even though, it is still much smaller than the $\mathcal O (2^n)$ cost of a brute-force classical simulation.  Despite the exponential asymptotic scaling of the sampling cost, the potential advantage of the method proposed in this work is given by its operational simplicity and, perhaps, by the practical convenience of just taking more shots instead of designing and running exponentially hard classical simulations. It is worth noting that standard validation methods for large-scale computations (close to the quantum supremacy regime) require advanced HPC classical simulations which are technically involved, economically expensive, and energetically inefficient. Therefore, the possibility of trading classical simulation cost for quantum sampling cost opens up a new and interesting perspective.

An important aspect to consider is that the number of collisions observed for a single device provides a good noise characterization benchmark only if most errors are incoherent and if the experiment is fair (non-advarsarial). Indeed, the benchmark could be intentionally or unintentionally spoofed by any mock device that samples from a sufficiently non-uniform distribution, even if unrelated to the correct quantum state. For the same reason, systematic coherent errors such as miscalibration of gates are undetectable by counting collisions on a single device,  since a wrong pure random state gives the same collision statistics as the correct pure random state. This implies that the number of observed collisions in a single quantum computer should not be used as a conclusive and absolute validation benchmark, but as a simple and practical diagnostic test augmenting the toolkit of existing benchmarks.
In Table \ref{tab:comparison} we summarize the main properties of collision-based benchmarks and we compare them with other randomized methods.

We also propose to use a similar collision-based protocol for the {\it cross-validation} of two independent quantum computers: by counting the number of cross-collisions observed when sampling two quantum computers running the same random circuit, one can estimate the joint quality of both devices. Similarly, one can use a reliable and trusted quantum computer to validate an untrustworthy device. Quite interestingly, compared to the single-device case, the cross-validation test between two independent quantum computers is much more difficult to spoof, as discussed in Sections \ref{sec:cross-collisions} and \ref{sec:conclusions}. Moreover, the cross-collision benchmark is also sensitive to coherent errors under the reasonable assumption that the noise models of the two quantum computers are statistically independent.

This article is organized as follows. In Sec.\ \ref{sec:collisions} we derive closed analytical formulas for the expected number of collisions when sampling from pure random states and when sampling from the uniform distribution. In Sec.\ \ref{sec:cv_test}, we introduce the collision volume benchmark. In Sec.\ \ref{sec:cross-collisions} we introduce the cross-collision volume benchmark, based on the observed collisions between two independent devices. In Sec.\ \ref{sec:collisions_with_noise}
we evaluate the relationship between the expected number of collisions and the noise level. Eventually, in \ref{sec:supremacy_regime} we give practical estimates of the sampling cost required by the validation approach proposed in this work.

\section{Expected number of collisions} \label{sec:collisions}

As a first step, we compute the expected number of collisions observed when sampling $N$ bitstrings of length $n$ from an arbitrary probability distribution $\{p_j: j=1 \dots D\}$ over $D=2^n$ possible outcomes.

Let $\hat I_j$ be an indicator random variable which is 1 if the outcome $j$ is sampled at least once and $0$ if $j$ is never sampled. The probability of the "$0$" event is $(1-p_j)^N$ and the probability of the "$1$" event is $1 - (1-p_j)^N$.
The number of distinct samples is $\hat W = \sum_{j=1}^D \hat I_j$, and so the number of collision events (i.e.~events in which a previously sampled bitstring is re-sampled) can be expressed as the random variable $\hat R=N - \hat W = N-\sum_{j=1}^D \hat I_j$. The expected number of collisions is:
\begin{equation} \label{eq:expected_collisions}
\mathbb{E}(\hat R) = N - \sum_{j=1}^D \mathbb{E}(\hat I_j)=N - D + \sum_{j=1}^{D}(1-p_j)^N.
\end{equation}
Equation \eqref{eq:expected_collisions} is exact and is valid for any probability distribution. For the particular case of the  uniform distribution $u=\{p_j=1/D, j=1 \dots D \}$ we get:
\begin{align} \label{eq:u_collisions}
\mathbb{E}_u(\hat R) &= N - D + D(1-1/D)^N  \simeq N - D(1 -e^{-N/D}),
\end{align}
where, in the last step, we used $(1-1/D)^D\simeq e^{-1}$ which is a good approximation for large $D$.
For $N/D \ll 1$, the expected collision frequency can be expanded as follows \cite{footnote2}:

\begin{align} \label{eq:collisions_u_approx}
\frac{\mathbb{E}_u(\hat R) }{N}&=\frac{N}{2D} + \mathcal O \left[\left(\frac{N}{D}\right)^2\right].
\end{align}

When sampling from a pure random quantum state $|\psi\rangle$, the probability distribution  $q=\{p_j: j=1 \dots D \}$ of the measurement outcomes is not uniform. For a typical Haar-random state $|\psi\rangle$, the number of probabilities $p_j$ within the interval $[p, p+dp]$ is well approximated by
$D P(p)dp$, where $P(p)=De^{-D p}$ is the Porter-Thomas distribution \cite{boixo2018characterizing, arute2019quantum}. So, transforming the sum in Eq.~\eqref{eq:expected_collisions} into an integral over the Porter-Thomas distribution, we get:
\begin{align} \label{eq:q_collisions}
\mathbb{E}_{q}(\hat R) &=N - D + D^2 \int_0^1  (1-p)^N e^{-D p}dp \nonumber  \\
&\simeq N - D + D^2 \int_0^\infty e^{-(N + D)p}dp, \nonumber \\
&= N - D +  \frac{D^2}{N + D}= \frac{N^2}{N + D},
\end{align}
where the approximation in the second line is valid for $D \gg 1$ since, in this regime, the term $e^{-Dp}$ exponentially suppresses the integrand for large values of $p$ such that the approximation $(1 - p)^{1/p} \simeq e^{-1}$ is justified and the upper integration limit can be extended to $+\infty$.
For $N/D \ll 1$, the expected collision frequency can be expanded as follows \cite{footnote2}:

\begin{align} \label{eq:q_collisions_approx}
\frac{\mathbb{E}_q(\hat R) }{N}&= \frac{N}{D} + \mathcal O \left[\left(\frac{N}{D}\right)^2\right].
\end{align}

In Fig.\ \ref{fig:collisions} we numerically simulate sampling experiments from the distribution of a random pure quantum state and from the uniform distribution, demonstrating a good agreement of the numerical results with the theory.
From a direct inspection of Fig.\ \ref{fig:collisions}  and of Eqs.\ (\ref{eq:u_collisions} 
- \ref{eq:q_collisions_approx}), we deduce the following facts:

\noindent (i) The expected number of collisions scales quadratically with the number of samples $N$ (up to higher order corrections).

\noindent (ii) To observe some collisions we need at least a number of samples of the order of $\sqrt{D}=2^{n/2}$. This is much smaller than the number of possible outcomes $D=2^n$, but it still scales exponentially with $n$. To observe a significant number of collisions ($\gtrsim 500$), we suggest the following empirical rule for the number of samples (see also the vertical line in Fig.\ \ref{fig:collisions}):
    \begin{equation}\label{eq:educated_guess}
    N \approx 32 \sqrt{D} = 2^{n/2 +5}.
    \end{equation}
Of course, $N$ can also be adaptively determined by continuously collecting measurements until enough collisions are observed.

\noindent (iii) The number of collisions can be used to distinguish the measurement distribution of a random pure state (e.g. the results of an ideal quantum computer) from the uniform distribution (e.g. the results of a very noisy quantum computer or of a classical pseudo-random number generator). In the limit $N/D\rightarrow 0$ the number of quantum collisions is twice the number of classical collisions. This fact is consistent with the similar doubling behavior of the linear XEB and of the collision probability \cite{footnote3},
see e.g.~Ref.~\cite{aaronson2019classical} and Theorems 2.10 and 3.5 of Ref.~\cite{liu2021moments}.
The gap is smaller for finite values of $N/D$, e.g., for $D=2^{16}$ and $N=32 \sqrt{D}\simeq 8.2 \times10^3$, we get $\mathbb{E}_{q}(\hat R)/\mathbb{E}_{u}(\hat R)\simeq 1.85$.

The previous observations can be operationally quantified by the {\it collision anomaly} parameter and the {\it collision volume} test that we define in the next section.


\section{A collision volume test}
\label{sec:cv_test}

\begin{figure}[!t]
    \includegraphics[width=1.0 \linewidth]{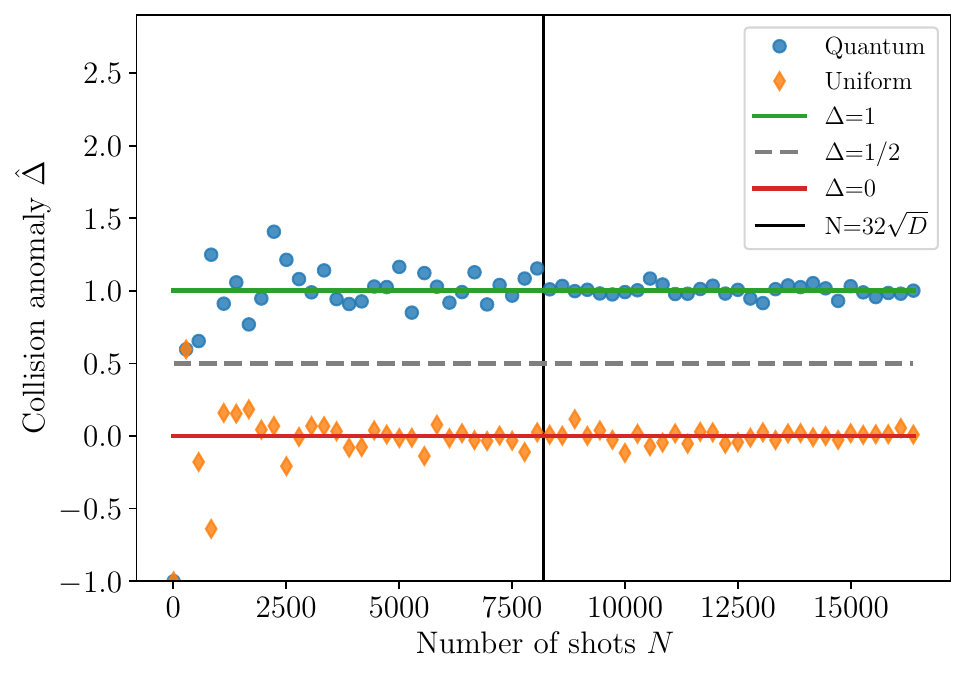}
    \caption{Numerical simulation of the quantum collision anomaly $\Delta$ defined in Eq. \eqref{eq:anomaly}, for a $16$-qubit random quantum state.} \label{fig:anomaly}
\end{figure}

With the aim of detecting a deviation of the ideal quantum distribution from the uniform distribution, we define the normalized {\it collision anomaly} as follows:
\begin{align}\label{eq:anomaly}
\hat \Delta &:= \frac{\hat R -\mathbb{E}_{u}(\hat R)}{\mathbb{E}_{q}(\hat R)- \mathbb{E}_{u}(\hat R)} = \frac{\hat R - N + D \left( 1-  e^{-N/D}\right)}{D^2/(N+D) -  D\, e^{-N/D}  },
\end{align}
which quantifies the deviation of the observed number of collisions $\hat R$ from the base value associated to the uniform distribution. By construction, the expected value of $\hat \Delta$ is $1$ when sampling from a typical random quantum state, while it is $0$ when sampling from the uniform distribution. See Fig.~\ref{fig:anomaly} for a numerical simulation.

Inspired by the well-established quantum volume test \cite{cross2019validating, blume2020volumetric, baldwin2022re}, we set $\hat \Delta \ge 1/2$ as a threshold for a sufficiently good quantum computation, and we define the following test based on the observed number of collisions, as schematized in Algorithm \ref{alg:collision_test}. 

\begin{algorithm}[H]
  \caption{Pseudo code for the quantum {collision volume (CV) test}.} \label{alg:collision_test}
   \begin{algorithmic}[1]
   \item $C \longleftarrow$ Generate an $n$-qubit random {\it quantum volume} circuit \cite{cross2019validating}
   \item $D \longleftarrow$ $2^n$
   \item $N \longleftarrow$ $32 \sqrt{D}$ 
   \item $\{b_j\} \longleftarrow$ Sample $N$ bitstrings by running $C$ with $N$ shots
   \item $\hat R \longleftarrow N$ - number of distinct bitstrings in $\{b_j\}$
   \item {\bf if} $\hat R < 500$: 
   \item $\qquad$ Increase $N$ (e.g. $N\longleftarrow 2 N$)
   \item $\qquad$ Go to line 4
   \item {\bf endif}
   \item $\hat \Delta \longleftarrow  \dfrac{\hat R - N + D \left( 1-  e^{-N/D}\right)}{D^2/(N+D) -  D\, e^{-N/D}  }$   \hspace{1 em} See Eq.\ \eqref{eq:anomaly}
   \item {\bf if} $\hat \Delta > 1/2$:
   \item $\qquad$ The collision volume test {\bf passed} for $n$ qubits
   \item {\bf else}:
   \item $\qquad$ The collision volume test {\bf failed} for $n$ qubits
   \end{algorithmic}
\end{algorithm}

\begin{definition}
We say that a device has a quantum {\it collision volume} (CV) equal to $2^n$ (or a log-CV of $n$), if it passes the CV test for $n$ qubits.
\end{definition}

\section{Cross-validation of two quantum computers}\label{sec:cross-collisions}

In this section, we are interested in the expected number of cross-collisions when sampling from two different probability distributions corresponding, e.g., to two different quantum computers operated by independent agents: Alice and Bob.
Let $\{p_j^{(A)}: j=1 \dots D\}$ and $\{p_j^{(B)}: j=1 \dots D\}$ be the probability distributions of Alice and Bob, respectively.
If Alice samples $N_A$ bitstrings, and Bob samples $N_B$ bitstrings, what is the expected number of cross-collisions? There can be slightly different ways of defining the number of cross-collisions.  Here we choose a definition that allows us to recycle most of the previous theory developed for a single device. 

Let $\hat R_{AB}$ be the number of global collisions in the merged dataset which contains all the bitstrings sampled by Alice and Bob.
The number of cross collisions can be defined by subtracting the number of local collisions from $\hat R_{AB}$:
\begin{equation}\label{eq:cross_collisions}
\hat R_X = \hat R_{AB} - \hat R_A - \hat R_B.
\end{equation}
We already know how to compute the expected values of $R_A$ and $R_B$, since they are given by Eq.\ \eqref{eq:expected_collisions}. Therefore,  what we need to compute here is the expected value of $\hat R_{AB}$.

Let the indicator $\hat I_j^{(AB)}$ be a random variable which is 1 if the outcome $j$ is sampled at least once by Alice or by Bob and $0$ if $j$ is never sampled. The probability of the "$0$" event is $(1-p^{(A)}_j)^{N_A} (1-p^{(B)}_j)^{N_B}$ and the probability of the "$1$" event is $1 - (1-p^{(A)}_j)^{N_A} (1-p^{(B)}_j)^{N_B}$.
The number of distinct samples is $\hat W_{AB}=\sum_{j=1}^D \hat I_j^{(AB)}$, and so the number of collision events can be expressed as the random variable $\hat R_{AB}=N_A + N_B -\hat W_{AB}$.
Its expected value is:
\begin{equation}\label{eq:merged_collisions}
\mathbb{E}(\hat R^{(AB)}) = N_A + N_B - D +  \sum_{j=1}^D (1-p^{(A)}_j)^{N_A} (1-p^{(B)}_j)^{N_B}.
\end{equation}

\begin{figure}[!tb] 
    \includegraphics[width=1.0 \linewidth]{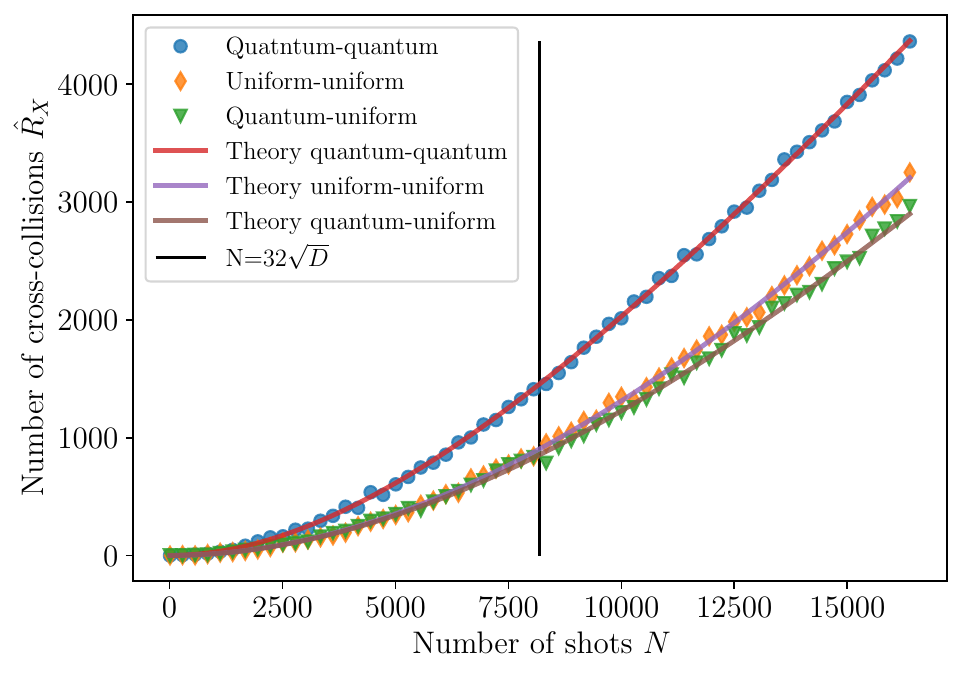}
    \caption{Numerical simulation of the number of cross-collisions $\hat R_X$ with respect to the number of samples $N$. We consider the cases in which Alice and Bob sample, with the same number of shots, from the same (16-qubit) random quantum states (blue), from the uniform distribution (orange), from a random quantum state and the uniform distribution (green). The theoretical curves correspond to Eqs. \eqref{eq:uu_collisions}, \eqref{eq:qq_collisions} and  \eqref{eq:qu_collisions}.} \label{fig:collisions_x}
\end{figure}

\begin{figure}[!tb]
    \includegraphics[width=1.0 \linewidth]{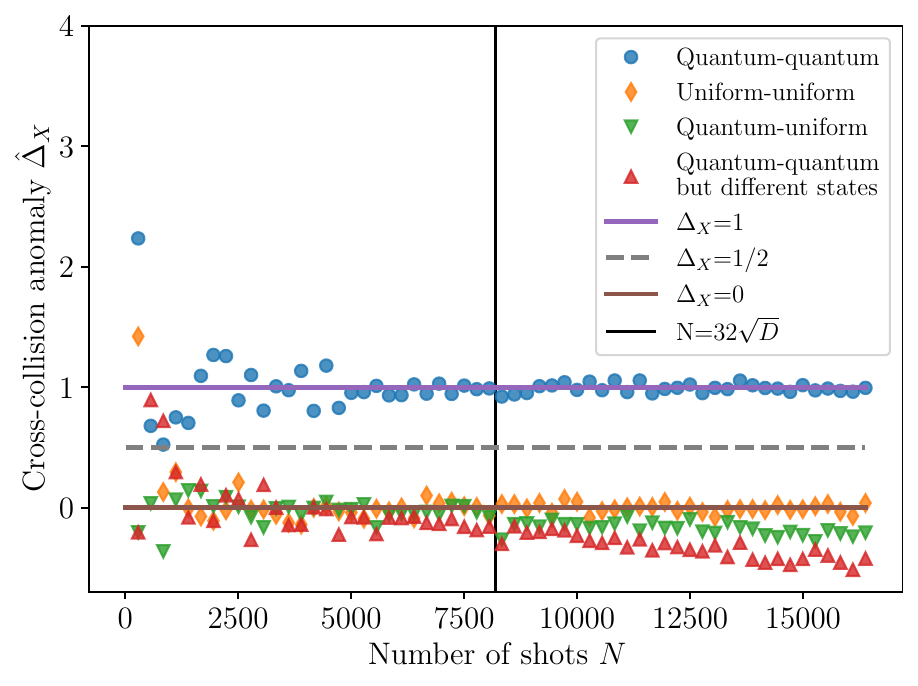}
    \caption{Numerical simulation of the cross-collision anomaly $\hat \Delta_X$ defined in Eq. \eqref{eq:anomaly}.
    We consider the cases in which Alice and Bob sample with the same number of shots from: the same 16-qubit random quantum states (blue), from the uniform distribution (orange), from a random quantum state and the uniform distribution (green), and from different random states (red).} \label{fig:anomaly_x}
\end{figure}
When Alice and Bob sample from the same distribution, Eq.\ \eqref{eq:merged_collisions} reduces to \eqref{eq:expected_collisions} with $N=N_A + N_B$, as expected.

If both Alice and Bob sample from the uniform distribution, using Eq.\ \eqref{eq:u_collisions}, we get the following expectation value for the number of cross-collisions:
\begin{equation} \label{eq:uu_collisions}
\mathbb E_{uu}(\hat R_X) = D [1 - e^{-N_A/D} - e^{-N_B/D} + e^{- (N_A + N_B)/D} ] .
\end{equation}
If both Alice and Bob sample from the distribution of the same pure random quantum state, using Eq.\ \eqref{eq:q_collisions}, we get:
\begin{equation} \label{eq:qq_collisions}
\mathbb E_{qq}(\hat R_X) = \frac{(N_A + N_B)^2}{N_A + N_B + D} -  \frac{N_A^2}{N_A + D} -  \frac{ N_B^2}{N_B + D} .
\end{equation}
Finally, if Alice samples from a pure quantum state and Bob samples from the uniform distribution, from Eq.\ \eqref{eq:merged_collisions} we get:
\begin{align}\label{eq:qu_collisions}
\mathbb E_{qu}(\hat R_X) &= N_A + N_B - D +\frac{ D^2  e^{-N_B/D} }{D+ N_A} -\mathbb E_{q}(\hat R_A) - \mathbb E_u(R_B)  \nonumber \\
&= \frac{N_A D}{N_A + D} (1 - e^{-N_B/D}),
\end{align}
where we applied the same derivation based on the Porter-Thomas distribution that we used to obtain Eq.\ \eqref{eq:q_collisions}.
A numerical analysis of the number of cross-collisions $\hat R_X$ is reported in Fig.\ \ref{fig:collisions_x}.

In the limit $D \rightarrow \infty$ (not always a good approximation), we have:
\begin{align}
\mathbb E_{qq}(\hat R_X) &\simeq  2 \frac{N_A N_B}{D} \nonumber \\ 
\mathbb E_{uu}(\hat R_X)  &\simeq  \mathbb E_{qu}(\hat R_X) \simeq \frac{N_A N_B}{D}  
\end{align}
So, in the limit $D \rightarrow \infty$, when both Alice and Bob sample from the same quantum state the number of cross-collisions is twice the value of the other cases involving the uniform distribution.
From numerical evidence (see e.g. Figs.~\ref{fig:collisions_x} and \ref{fig:anomaly_x}) and from intuitive arguments \cite{footnote4}, we conjecture that $\mathbb E_{qq'} (\hat R_X) \le \mathbb E_{uu} (\hat R_X) < \mathbb E_{qq}(\hat R_X)$, where $q$ and $q'$ are associated to different pure random states. In other words, Alice and Bob get more cross-collisions if they both sample from the same quantum state with sufficiently good fidelity. In this sense, the observation of a large number of collisions is a good validation test for the joint quality of both devices. Moreover, if Alice has a sufficiently good and trusted quantum computer, she can use it as a reference for testing the quality of an untrusted quantum computer (Bob's device), by simply measuring the number of cross-collisions without performing any classical simulation.

The previous observations suggest to define the following normalized cross-collision anomaly:

\begin{equation}
\hat \Delta_X = \frac{\hat R_X - \mathbb E_{uu}(\hat R_X)}{\mathbb E_{qq}(\hat R_X) - \mathbb E_{uu}(\hat R_X)}.
\end{equation}

 By construction, the expected value of $\hat \Delta_X$ is $1$ when both Alice and Bob are sampling from the same random pure quantum state while it is $\lesssim 0$ if Alice or Bob (or both) are sampling from the uniform distribution.
Numerical simulations of the cross-collision anomaly $\Delta_X$ are reported in Fig.\ \ref{fig:anomaly_x}.

Similarly to the single-device case, we define the following {\it cross-collision volume} (XCV) test, as schematized in Algorithm \ref{alg:x_collision_test}.

\begin{algorithm}[H]
  \caption{Pseudo code for the  quantum {\it cross-collision volume  (XCV) test.}}\label{alg:x_collision_test}
   \begin{algorithmic}[1]
   \item $C \longleftarrow$ Generate an $n$-qubit random {\it quantum volume}  circuit \cite{cross2019validating}
   \item $D \longleftarrow 2^n$
   \item $N \longleftarrow 32 \sqrt{D}$
   \item $\lambda \longleftarrow 1$ (Optionally, use a different shot ratio)
   \item $N_A \longleftarrow \lambda \; N$
\item $N_B \longleftarrow \lambda^{-1}\; N$ 
   \item $\{b_j^{(A)}\} \longleftarrow$ Run $C$ on Alice's device with $N_A$ shots
    \item $\{b_j^{(B)}\} \longleftarrow$ Run $C$ on Bob's device with $N_B$ shots

    \item $\hat W_{AB} \longleftarrow $ Number of distinct bitstrings in $\{b^{(A)}_j\} \cup \{b^{(B)}_j\}$
    
    \item $\hat W_A \longleftarrow $ Number of distinct bitstrings in $\{b^{(A)}_j\}$

   \item $\hat W_B \longleftarrow $ Number of distinct bitstrings in $\{b^{(B)}_j\}$

   \item $\hat R_X \longleftarrow \hat W_A + \hat W_B- \hat W_{AB} $ (Equivalent to Eq. \eqref{eq:cross_collisions})

   \item {\bf if} $\hat R_X < 500$: 
   \item $\qquad$ Increase $N$ (e.g. $N\longleftarrow 2 N$)
   \item $\qquad$ Go to line 5
   \item {\bf endif}
   \item $\hat \Delta_X \longleftarrow \dfrac{\hat R_X - \mathbb E_{uu}(\hat R_X)}{\mathbb E_{qq}(\hat R_X) - \mathbb E_{uu}(\hat R_X)}$ (using Eqs. \eqref{eq:uu_collisions} and \eqref{eq:qq_collisions})
   \item {\bf if} $\hat \Delta_X > 1/2$:
   \item $\qquad$ The cross-collision volume test {\bf passed} for $n$ qubits
   \item {\bf else}:
   \item $\qquad$ The cross-collision volume test {\bf failed} for $n$ qubits
   \end{algorithmic}
\end{algorithm}

\begin{definition}
    We say that two devices, A and B, have a {\it cross-collision volume} (XCV) equal to $2^n$ (or a log-XCV equal to $n$), if they pass the XCV test for $n$ qubits.
\end{definition}

It is useful to contextualize our cross-collision test within the more general problem of cross-platform verification \cite{greganti2021cross, linke2017experimental, elben2020cross}. In Ref.\ \cite{elben2020cross} a method for estimating the fidelity between the output states generated by two different quantum computers was proposed. Compared to \cite{elben2020cross}, our protocol is conceptually and technically simpler since it does not require measuring different Pauli operators. On the other hand, the method of  Ref.\ \cite{elben2020cross} can be used to evaluate the fidelity between states prepared by arbitrary circuits, while our protocol is designed for the specific case of random quantum circuits. 
A similar comparison applies with respect to other cross-platform consistency tests, which are based on correlations between the results of specific quantum algorithms or specific observables  \cite{linke2017experimental, elben2020cross}. 

 \section{Expected number of collisions for a quantum device subject to depolarizing noise}
\label{sec:collisions_with_noise}

In the previous sections, we studied the expected number of collisions involving pure states, i.e., the results of a noiseless quantum computer. Unfortunately, all existing quantum computers are noisy. What happens to the expected number of collisions if a quantum computer is noisy?

It has been numerically and experimentally demonstrated \cite{boixo2018characterizing, arute2019quantum} (see also \cite{dalzell2024random} for theoretical arguments), that the quantum state generated by a noisy quantum computer after the execution of a random circuit is typically  well approximated by the following density matrix:
\begin{equation}\label{eq:depolarized}
\rho_\alpha = \alpha |\psi\rangle \langle \psi | + (1 - \alpha) \mathbb{I}/D,
\end{equation}
where $|\psi\rangle$ is the pure random state that would be generated without noise and $\mathbb{I}/D$ is the maximally mixed state. For $D \gg 1 $, $\alpha \in [0, 1]$ is approximately equal to the quantum state fidelity between the noisy state $\rho_\alpha$ and the ideal state $|\psi\rangle$, {\it i.e.}, $\alpha\approx \langle \psi| \rho_\alpha |\psi \rangle$.

\begin{figure}[!tb] 
    \includegraphics[width=1.0 \linewidth]{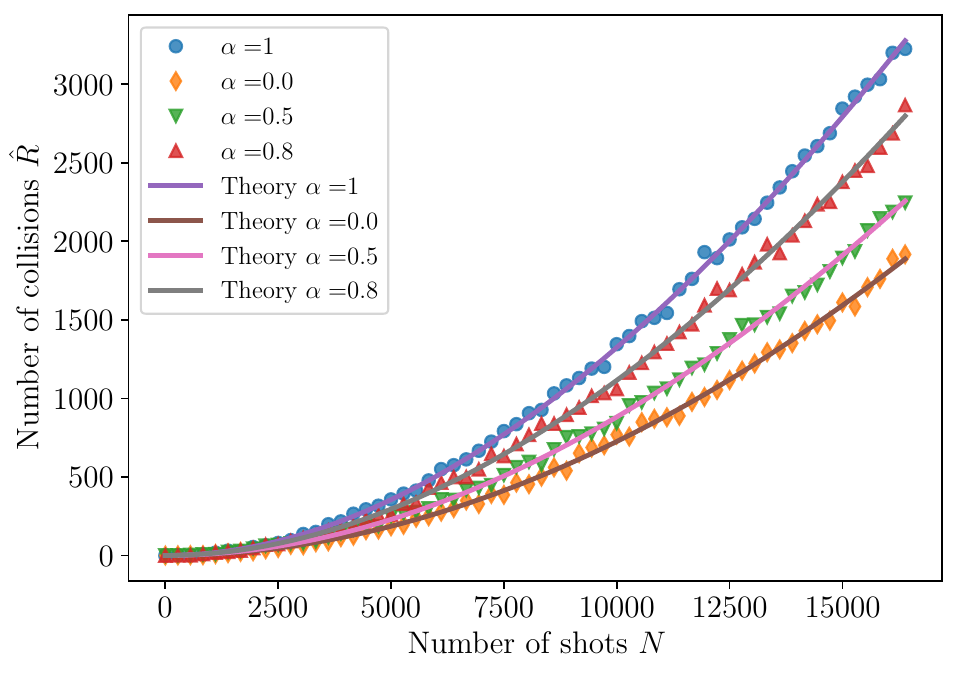}
    \caption{Numerical simulation of the number of collisions $\hat R$ with respect to the number of samples $N$,  for a $16$-qubit noisy random state for different values of fidelity $\alpha$ as defined in Eq.\ \eqref{eq:depolarized}. 
    The theoretical curves correspond to Eq.\ \eqref{eq:q_collisions_noisy}.
    Note that for $\alpha=1$ and $\alpha=0$, we recover the results plotted in Fig.\ \ref{fig:collisions}. }
    \label{fig:collisions_alphas}
\end{figure}

The effect of depolarizing noise described in \eqref{eq:depolarized} is to deform the distribution of the bitstring probabilities in a continuous way (see e.g.\ \cite{boixo2018characterizing}): from the Porter-Thomas limit $P_{\alpha=1}(p)=De^{-Np}$ towards the uniform limit $P_{\alpha=0}(P)=\delta(p - 1/D)$.  To estimate the number of collisions, we could repeat all the previous calculations replacing the ideal Porter-Thomas distribution with the deformed distribution.
However, we will use a simpler approach based on the probabilistic interpretation of Eq.\ \eqref{eq:depolarized}.

\begin{figure}[!tb]
    \includegraphics[width=1.0 \linewidth]{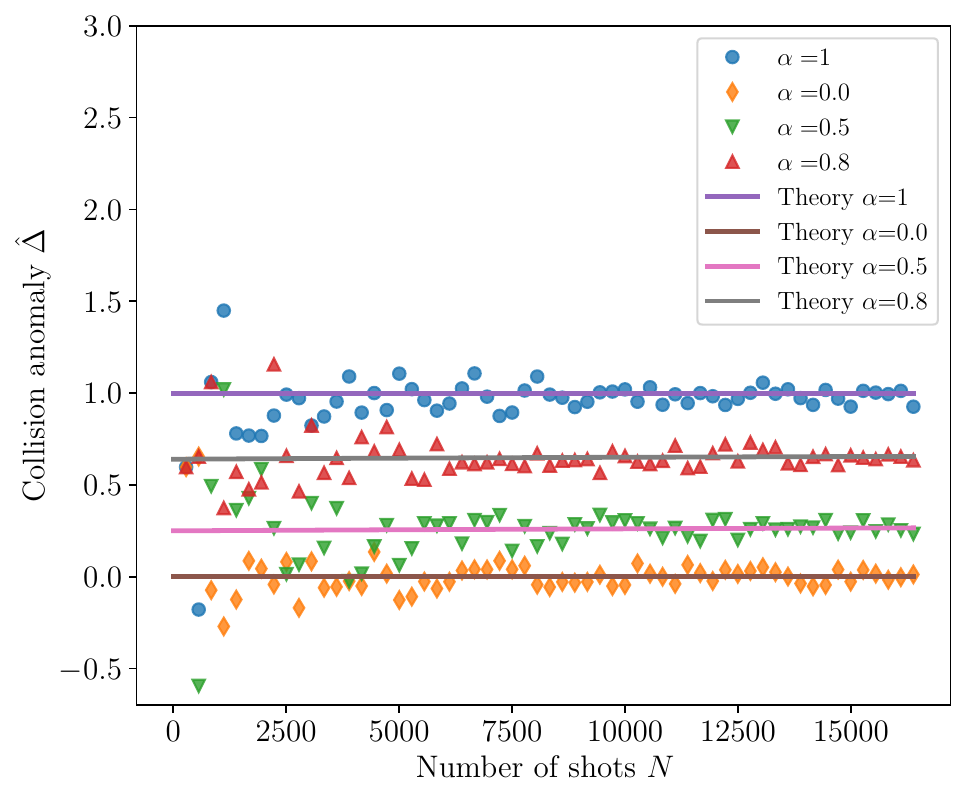}
    \caption{Numerical simulation of the quantum collision anomaly $\hat \Delta$ defined in Eq. \eqref{eq:anomaly}, for a $16$-qubit random quantum state subject to global depolarizing noise with fidelity parameter $\alpha$ as defined in Eq.\ \eqref{eq:depolarized}.} \label{fig:anomaly_alphas}
\end{figure}

Assume that we have two fictitious agents, Alice who measures with a perfect quantum computer $N_A= \alpha N$ bitstrings from $|\psi\rangle$ and Bob who measures $N_B=N(1-\alpha)$ bitstrings from the uniform distribution.  In the limit of large N, if they merge and randomly shuffle their results, they effectively obtain $N$ samples from the noisy quantum state in Eq.\ \eqref{eq:depolarized}. Moreover, if we are just interested in counting the number of collisions, the shuffling operation is irrelevant and we can consider the bitstrings of Alice and Bob as distinct datasets. But this is exactly the situation that we have studied in the previous section, where we derived Eq.\ \eqref{eq:merged_collisions}. More explicitly, if $q_\alpha$ is the  probability distribution associated with the noisy quantum state  \eqref{eq:depolarized}, the expected number of collisions is:
\begin{equation}\label{eq:q_collisions_noisy}
\mathbb{E}_{q_\alpha}( \hat R) =  \mathbb{E}_{qu}(\hat R^{AB})\Big|_{\substack{N_A=\alpha N \ \ \ \ \\ N_B = (1- \alpha) N}}= N - D + \frac{D^2  e^{- (1- \alpha) N/D}}{\alpha N + D}.
\end{equation}
Numerical simulations of the number of collisions for different noise levels are shown in Fig.\ \ref{fig:collisions_alphas}.
In the limit $N/D \ll 1$, the first-order approximation of the collision frequency is
\begin{equation}
\frac{\mathbb{E}_{q_\alpha}( \hat R)}{N}= (\alpha^2 + 1) \frac{N}{2D} + \mathcal O \left[\left(\frac{N}{D}\right)^2\right],
\end{equation}
corresponding to a constant floor of $0.5 N/D$ (that of the uniform distribution)  plus an additive quantum signal scaling as $\alpha^2$ which is the deviation that we are interested in measuring. 
Assuming that statistical fluctuations depend weakly on $\alpha$, we can counteract the effect of noise by scaling $N$ as $N \rightarrow \alpha^{-1} N$ in order to obtain a signal comparable to the noiseless case. Accordingly, we update the empirical educated guess given in Eq.\ \eqref{eq:educated_guess} with a new formula that takes noise into account:
\begin{equation}\label{eq:educated_guess_noisy}
N \approx \frac{32 \sqrt{D}}{\alpha} = \frac{2^{n/2+5}}{\alpha}.
\end{equation}

More rigorously, from Eq.\ \eqref{eq:q_collisions_noisy} we can compute the expected value of the collision anomaly defined in Eq.\ \eqref{eq:anomaly}, obtaining the following function of $\alpha, N$ and $D$:

\begin{align}\label{eq:noisy_anomaly}
\mathbb E_{q_\alpha} (\hat \Delta) =& \left[\frac{e^{- (1- \alpha) N/D}}{\alpha N + D} - \frac{e^{- N/D}}{D}\right] / \left[\frac{1}{N + D} - \frac{ e^{- N/D}}{D}\right].
\end{align}
As expected, $\mathbb E_{q_\alpha}(\hat \Delta)$ is a monotonically increasing function of  $\alpha$, interpolating from 0 (at $\alpha=0$) to 1 (for  $\alpha=1$).  Numerical simulations of the collision anomaly for different noise levels are shown in Fig.\ \ref{fig:anomaly_alphas}.
In the limit $N/D \ll 1$, we have
\begin{align}\label{eq:noisy_anomaly_approx}
\mathbb E_{q_\alpha} (\hat \Delta) &=\alpha^2 +  \mathcal O\left(\frac{N}{D}\right).
\end{align}
Hence, given a measurement of the collision anomaly one can estimate the fidelity $\alpha$ of the quantum computation either by taking the square root (first order approximation), or by solving the more accurate theoretical expression in \eqref{eq:noisy_anomaly} for $\alpha$.

\section{Sampling cost: from the collision volume regime up to the quantum supremacy regime}
\label{sec:supremacy_regime}

How many shots are necessary to run a collision volume test or a cross-collision volume test for, say $n=20$ qubits? How many shots are necessary to observe a small but non-zero collision anomaly in a quantum supremacy experiment with, say, $n=53$ qubits?

\begin{figure}[!tb]
    \includegraphics[width=1.0 \linewidth]{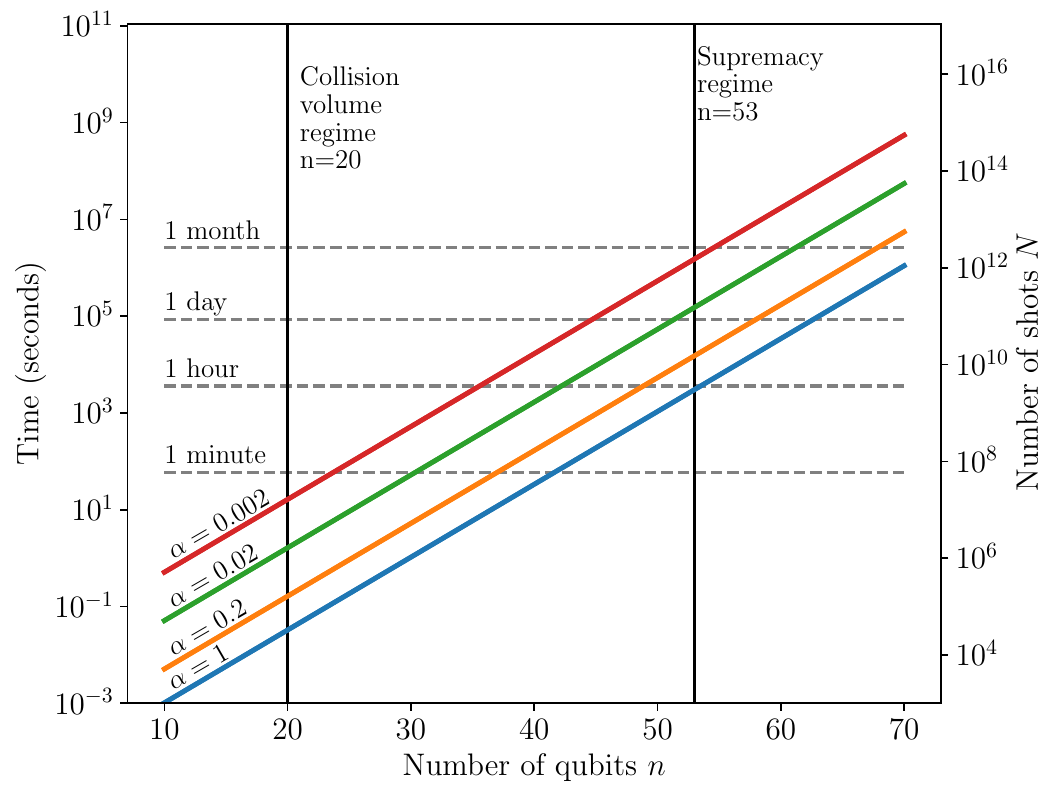} 
    \caption{Execution time (left axis) and required number of shots (right axis) necessary to observe a clear collision anomaly for different values of the fidelity $\alpha$. The number of shots is deduced from Eq.\ \eqref{eq:educated_guess_noisy}. The execution time is estimated from $N$ assuming a repetition rate of 1 million shots per second.}
    \label{fig:cost}
\end{figure}

In Fig.\ \ref{fig:cost} we give a cost estimate in terms of the number of shots and in terms of the execution time in both regimes: the collision volume regime (few qubits) and the quantum supremacy regime (many qubits).
We observe that running a collision volume test for state-of-the-art NISQ computers ($n\approx 20$ effective qubits with $\alpha > 0.5$) is quite easy since $N\le 10^5$ shots are enough.
On the other hand, for validating quantum computations in the quantum supremacy regime ($n>53$ and $\alpha < 0.002$) by empirically estimating  $\hat \Delta\simeq \alpha^2 >0$ or $\hat \Delta_X \simeq \alpha^2 > 0$ with sufficient statistical confidence, at least $N>10^{12}$ shots are required. Assuming a measurement repetition rate of 1 MHz, this corresponds to more than 3 weeks of sampling time (see vertical line in Fig.\ \ref{fig:cost}).
Running a weeks-long quantum experiment is arguably infeasible with current technology, but it may become feasible with future technology.

For example, with $M$ independent processors (e.g.\ in a multi-core QPU), one can reduce the total time by a factor of $1/M$. Moreover, it is reasonable to expect fidelity improvements, from $\alpha\approx 0.002$ as estimated in Ref.\ \cite{arute2019quantum} to larger values. Depending on the magnitude of such technological improvements, the large quantum sampling cost required to validate quantum computations by measuring quantum collisions could become competitive with respect to the intensive classical simulation cost required by classical validation strategies  \cite{boixo2018characterizing, arute2019quantum}. The potential advantage of the collision-based validation method is perhaps even stronger if, instead of the time cost, we consider the energy consumption. Indeed, the experimental task quantum circuit sampling, even if technically demanding, can be more energy efficient than running large classical simulations on supercomputers \cite{scholes2010green, villalonga2020establishing, auffeves2022quantum}.

\section{Conclusions and outlook} \label{sec:conclusions}

In the context of random circuit sampling, we have shown that it is possible to estimate the quality of a quantum device by simply counting the number of observed collisions in the measurement outcomes.
Specifically, we introduced the notion of quantum collision volume (CV) and quantum cross-collision volume (XCV). The CV is a benchmark for a single quantum computer, while the XCV quantifies the joint performance of two quantum computers, by essentially testing whether their measurement outcomes are consistent with the same random quantum state. 
We also analytically estimated the way in which depolarizing noise decreases the expected number of collisions, showing how the state preparation fidelity can be empirically deduced from the {\it collision anomaly} observed in the measurement outcomes (see Eqs.~\eqref{eq:anomaly} and \eqref{eq:noisy_anomaly_approx}), without knowing the wave function of the ideal quantum state.

From the resource analysis presented in Sec.\ \ref{sec:supremacy_regime}, one can envisage that the approach presented in this work can be applied in two different contexts: (i) for benchmarking small-scale quantum computers and (ii) for the validation (or cross-validation) of large-scale quantum computations, avoiding intensive classical computing costs and the associated energy consumption. Due to the exponential scaling of the sampling cost, the second application is not currently feasible for circuits that are hard to classically simulate ($n>50$ qubits), but may become feasible in the future if the sampling rate and/or the state fidelity will increase by one or two orders of magnitude. The first application is instead immediately feasible and can be easily used to benchmark existing quantum processors.

From a theoretical point of view, the analysis introduced in this work can be further extended and deepened in many directions. For example, we highlight some open questions and open problems:

\begin{enumerate}
\item The collision volume (CV) test, if applied on a single device, can be easily deceived by a ``dishonest'' machine. For example, a trivial classical device that always outputs the same bitstring would artificially pass the CV test. The cross-collision volume (XCV) test defined in Sec.~\ref{sec:cross-collisions}
seems much more difficult to spoof, unless both Alice and Bob agree on a cheating strategy. Is it possible to give a rigorous proof of this fact? Can Alice rigorously certify, by running a cross-collision test, that Bob has a good quantum computer without trusting him?

\item How sensitive are collision benchmarks to different types of errors? For example, systematic coherent errors such as miscalibration of gates are undetectable by counting collisions on a single device,  since a wrong pure random state gives the same collision statistics as the correct pure random state. On the other hand, coherent errors are detectable by the cross-collision test applied on two independent devices, since different random pure states yield a small cross-collision anomaly (see  Fig.~\ref{fig:anomaly_x}). This fact suggests that, by comparing the number of local and cross-collisions, one may even deduce the relative impact between coherent and incoherent errors. 

\item In this work we assumed the possibility of preparing Haar-random quantum states. In practice, however, one can only run random circuits of polynomial depth, which are known to generate good but not perfect approximations of Haar-random states. What happens to the results derived in this work if we consider feasible random circuits instead of ideal Haar-random states? On a similar vein,
can we safely replace the Haar-random distribution with a unitary $t$-design?\\

\item Is it possible to apply similar benchmarks to random states that are subject to constraints, e.g. symmetries, in order to have a more peaked output distribution and therefore a larger quantum collision anomaly? Taking this question to the extreme limit, one can consider the recently proposed class of randomized circuits having the output distribution highly peaked on a specific random bitstring \cite{aaronson2024verifiable}. What is the physical meaning of collisions and cross-collisions in this case? For a small number of samples, most collisions will trivially come from the peak bitstring, but what happens if we increase the number of shots? \label{item:peak}

\item The validation approach based on quantum collisions   
 has a (quantum) sampling cost of the order of $2^{n/2}$ and zero classical computing cost. Beyond practical implementation aspects, is there a theoretical asymptotic advantage in the total number of (quantum and classical) operations, compared to other benchmarking methods such as QV  \cite{cross2019validating, blume2020volumetric, baldwin2022re} or XEB \cite{boixo2018characterizing, aaronson2019classical}?

\item Is there any workaround to reduce the high sampling cost of collision benchmarks? This seems an unavoidable limitation due to the small collision probability. Symmetries and more structured circuits (see point \ref{item:peak}) may help reduce the sampling cost but, at the same time, could make classical spoofing easier. Moreover, a brute-force way of reducing the wall-clock time of the experiment is exploiting multi-core quantum processors as discussed in Sec.~\ref{sec:supremacy_regime}. 

\item Can one extend our benchmarks beyond quantum computers that are based on the circuit model? For example, one could imagine benchmarking analog quantum devices, quantum simulators, continuous-variable photonic processors, etc., by running a protocol similar to Algorithm \ref{alg:collision_test} or   \ref{alg:x_collision_test} but where the random circuit $C$ is replaced by some
suitable state preparation procedure that approximately generates a Haar-random state.\\

\item In this work we always assumed a register of $n$ qubits. Can one extend the same results to $n$ qudits? Since we only used properties of Haar-random states in a global Hilbert space, we envisage that all results could be straightforwardly extended to qudits, up to minor adjustments, e.g., $D:=2^n \rightarrow d^n$, and bitstrings $\rightarrow$ ditstrings. 

\end{enumerate}

The above open questions are worth being investigated in future research.
We hope that this work will stimulate further theoretical investigations and that collision-based protocols will be adopted as practical experimental benchmarks for existing and future quantum computers.

\section{CODE AND DATA AVAILABILITY}
The code to reproduce all the numerical results reported in this work is available, in the form of a self-contained Jupyter notebook, at \href{https://github.com/unitaryfund/research/}{https://github.com/unitaryfund/research/}.

\section{Acknowledgments}
We thank the organizers and the participants of the NISQAH-2023 workshop in which some aspects of this work have been informally shared and discussed. In particular, we thank fruitful discussions with Scott Aaronson, who suggested to estimate the sampling cost in the quantum supremacy regime. We thank Nathan Shammah, William Zeng and Vincent Russo for useful comments and suggestions on this manuscript.  
We acknowledge support from Unitary Fund and from the PNRR MUR project PE0000023-NQSTI (Italy).

\bibliography{refs}

\begin{thebibliography}{10}
\bibitem{footnote1}
There can be different ways of defining the number of collisions. Given $k$
  equal samples, one can count all the $k (k-1)/2$ equal pairs, or the number
  of $k-1$ re-sampling events. In this work, we use the latter approach. In any
  case, both definitions become equivalent in the rare collision limit $k \le
  2$.
\bibitem{footnote2}
Neglecting higher-order terms in $N/D$ is not always a good approximation. We
  recommend avoiding the first-order approximation when looking for an accurate
  estimate.
\bibitem{footnote3}
In general, the expected number of collisions is not directly proportional to
  the collision probability. From Eq.~\eqref{eq:expected_collisions}, it is
  evident that the former depends on all the moments $\sum_j p_j^k$ of the
  probability distribution, while the latter corresponds to $\sum_j p_j^2$.
\end{thebibliography}
\bibliographystyle{unsrt}


\end{document}